\documentclass[preprint,showpacs,preprintnumbers,superscriptaddress]{revtex4-1}

\usepackage{multirow}
\usepackage{diagbox}
\usepackage{booktabs}
\usepackage{array}
\usepackage{natbib}
\usepackage{amsmath,amssymb,graphicx,bm,subfigure,float,siunitx,color}
\usepackage{bm,subfigure,float,siunitx,graphicx}
\usepackage[figuresright]{rotating}
\usepackage{color}
\usepackage{epstopdf}
\usepackage{mathrsfs}
\newcommand{\be}{\begin{equation}}
\newcommand{\ee}{\end{equation}}

\newcommand{\1}{\left}
\newcommand{\2}{\right}
\def\({\left(}
\def\){\right)}
\def\[{\left[}
\def\]{\right]}

\newcommand{\dif}{\,\mathrm{d}}

\newcommand{\al}{\alpha}

\newcommand{\sig}{\sigma}

\renewcommand{\th}{\theta}

\begin{document}
\title{\boldmath Thin accretion disk around black hole in
Einstein-Maxwell-scalar theory}
\author{Yingdong Wu }
\email{Email address: yingdongwu7@gmail.com}
\affiliation{Department of Physics, Southern University of Science and
Technology, Shenzhen 518055, Guangdong, China}

\author{Haiyuan Feng \footnote{Corresponding author}}
\email{Email address:  fenghaiyuanphysics@gmail.com}
\affiliation{Department of Physics, Southern University of Science and
Technology, Shenzhen 518055, Guangdong, China}

\author{Wei-Qiang Chen\footnote{Corresponding author}}
\email{Email address: chenwq@sustech.edu.cn}
\affiliation{Department of Physics, Southern University of Science and Technology, Shenzhen 518055, Guangdong, China}

\begin{abstract}

We examine the accretion process in a thin disk surrounding a supermassive black hole within the framework of Einstein-Maxwell-scalar (EMS) gravity. Our investigation aims to elucidate how variations in model parameters affect different physical properties of the disk. When keeping EMS parameters $\beta$ and $q$ constant, we observe a reduction in radiation flux and temperature as $\alpha$ increases. However, the luminosity and radiative efficiency exhibit relatively minor variation. Conversely, under fixed $\alpha$ and $q$, an escalation in $\beta$ leads to heightened levels of radiation flux, temperature, luminosity, and radiative efficiency. These results underscore the diverse influences of model parameters on observable metrics, providing valuable insights for the astronomical study of distinct black holes.

\end{abstract}

\maketitle

\section{Introduction}
A black hole emits no electromagnetic radiation and is solely detectable through its gravitational interactions with surrounding objects. Observations of stellar-mass black holes are feasible in select scenarios. Notably, the detection of gravitational waves resulting from the merger of binary black hole systems stands as compelling evidence for their existence \cite{50}. Moreover, stellar-mass black holes can be identified via X-ray emissions in binary systems. In such configurations, a visible companion star contributes matter to the black hole, forming an accretion disk, which emits observable X-ray radiation from its inner regions. In non-interacting binaries, the presence of a black hole can be deduced through radial velocity measurements of the companion star \cite{511}.

The initial investigation into accretion disks, employing a Newtonian framework, was conducted in Ref.\cite{51}. Subsequently, a series of papers by Novikov, Thorne, and Page explored the general relativistic approach \cite{52, 53, 54}. The model assumes a constant mass accretion rate independent of the disk's radius, implying a steady state. Additionally, it posits that accreting matter moves in Keplerian orbits, suggesting the absence of a strong magnetic field surrounding the central object. Furthermore, radiation emitted from the disk is modeled as black body radiation, resulting from the thermodynamic equilibrium within the disk.

The investigation of thin accretion disks encircling black holes presents a promising avenue for probing distinctions between General Relativity (GR) and alternative gravitational theories. Studies focusing on thin accretion disk models and their characteristics within modified gravity frameworks have been documented in several works \cite{55, 56, 57, 47, 48, 49}. In contexts involving additional spatial dimensions, such as Kaluza-Klein and Brane world scenarios, the behavior of accretion disks has been explored \cite{ 58, 59, 60}. Furthermore, the properties of accretion disks in theories like Chern-Simons and scalar-tensor-vector gravity have been addressed in the literature \cite{ 61, 62}. Investigating the dynamics of accretion disks around exotic objects such as wormholes has garnered interest, with relevant studies available \cite{63, 64, 65}.

Accretion disk phenomena have also been examined in relation to various compact astrophysical entities including neutron stars, boson stars, fermion stars, and gravastars, as evidenced by studies referenced in the literature \cite{66, 67, 68, 69, 70, 71, 72, 73, 74,106}. Notably, recent research \cite{75} delves into the properties of thin accretion disks surrounding electrically and magnetically charged Gibbons-Maeda-Garfinkle-Horowitz-Strominger (GMGHS) black holes. Additionally, techniques such as $k_{\alpha}$ iron line analysis and continuum-fitting methods have been employed to discern diverse astrophysical objects based on the characteristics of their accretion disks \cite{ 76, 78, 79}.

Additionally, the Einstein-Maxwell-scalar (EMS) theory serves as a simplified model extensively explored within the realms of gravity. Unlike conventional holographic superconducting or superfluid models \cite{ 25, 26, 27}, wherein the charged scalar field exhibits minimal coupling with the Maxwell field, EMS models feature a neutral scalar field interacting with the Maxwell field through a non-minimal coupling function \cite{ 28, 30,31,32,33,34,35,36,37,38,39,40,41,42,43,44, 46}. Categorization of EMS models into two distinct types, denoted as IIA and IIB, is contingent upon the nature of the coupling function and its ability to support Reissner-Nordstrm (RN) black holes in electrovacuum \cite{ 29}. This study focuses on static spherically symmetric EMS black holes, investigating the characteristics of thin accretion disks encircling them.

The organization of the paper is as follows. In section II, we give a brief description of the accretion process in EMS theory. In section III, we review the basic results of the geodesic equation and portray the ISCO of different parameters. In section IV, through the properties of the accretion disks in the EMS black hole, we discuss the radiant energy flux, the radiation temperature, the observed luminosity, and the Novikov-Thorne efficiency and some important values. Finally, Section V concludes the paper with a summary and an outlook.

\section{The black hole solution in Einstein-Maxwell-scalar theory}
	
The theory containing coupling between the scalar field and
gravity was first considered by Fisher who found a static and spherically
symmetric
	solution of the Einstein massless scalar field equations \cite{102}. It is
	known 	that
	the Einstein-Maxwell-scalar theory can emerge naturally in physics, for
	example, in the contexts of Kaluza-Klein models \cite{103},
	supergravity/string
	theory \cite{104} and cosmology \cite{105}.

  In EMS theory, the coupling between the scalar field and
    the
	electromagnetic
	field can be represented through interaction terms in the Lagrangian. Its
	primary origin can be traced back to the low-energy effective action of
	string theory, where the scalar field arises from the scalar modes of
	string theory, while the electromagnetic field originates from the gauge field
	interactions. The scalar field can also be identified as the dilaton field,
	which arises from compactification processes or the moduli space of the
	theory. The coupling between the scalar and electromagnetic fields is a
	result of the higher-dimensional theory compactified into four dimensions,
	where the scalar field acts as a dynamical variable that affects the
	coupling strength of the gauge fields. This coupling not only modifies the
	behavior of electromagnetic fields but also influences the solutions of the
	Einstein equations in the presence of the scalar field.
	These coupling terms are typically introduced to generate outcomes that
	align with
	experimental or astrophysical observations in specific scenarios. For
	instance, one common coupling term takes the form $\mathcal{L}_{\text{int}}
	= f(\phi)F_{\mu\nu}F^{\mu\nu}$, where $f(\phi)$ is a function of the
	scalar field $\phi$, and $F_{\mu\nu}$ is the electromagnetic field
	strength tensor. Depending on the form of $f(\phi)$, this
	coupling can be constant, exponential, or take other forms. Such
	interaction
	terms can modify the behavior of electromagnetic field propagation,
	especially in regions with a strong scalar field background, such as near
	black holes or on cosmological scales.

	Additionally, the introduction of a scalar field can also be seen
as a
	modification of the
	standard Einstein gravity. Such modified models are sometimes
	referred to as "scalar-tensor theories" or extended gravity theories. In
	these theories, the coupling between the scalar field and the
	electromagnetic field can alter the behavior of the electromagnetic field
	and change the solutions of the gravitational field. For instance, studies
	in EMS theory have shown that the presence of a scalar field can affect the
	electromagnetic structure of black holes, leading to new types of black
	hole solutions, such as scalar-electromagnetic coupled black holes.

We consider the  Einstein-Maxwell-scalar(EMS) theory described by the following action
\begin{equation}
	\label{1}
	S[g_{\mu\nu},\phi]=\frac{1}{16\pi}\int
	d^{4}x\sqrt{-g}(R-2\nabla_{\mu}\phi\nabla^{\mu}\phi-K(\phi)F_{\mu\nu}F^{\mu\nu}-V(\phi)),
\end{equation}
where $R$ is the Ricci scalar, $\phi$ is a real scalar field and $F^{\mu\nu}$ is the Maxwell field strength tensor. The EMS theory is the model that describes the real scalar field $\phi$ minimally coupled to GR and non-minimally coupled to Maxwell's background. $K(\phi)$ is the coupling function between the Maxwell and scalar field, and $V(\phi)$ is the scalar potential. if we set up the determined expressions of $K(\phi)$ and $V(\phi)$ first,
then the corresponding black hole solutions are determined.
 For example, if we set $K=1$ and $V=2\lambda$, $\lambda$ is the cosmological
constant, then the EMS theory gives the Reissner-Nordstrom-de Sitter solution. And if we set $K=e^{2\phi}$ and $V=0$, then we can get the dilation black hole solution \cite{1,2}.  Some important solutions with different $K(\phi)$
and $V(\phi)$ are given by \cite{3,4,5,6,7,8}.


In our study, we consider V=0 and the coupling function $K(\phi)$ is (see Appendix A for more details)
\be
\label{5}
K(\phi)=\frac{(\al^{2}+1)e^{\frac{-2\phi}{\al}}}{(\al^{2}+1+\beta)e^{\frac{-2\phi(\al^{2}+1)}{\al}}+\beta
 \al^{2}},
\ee
with $\alpha$ and $\beta$ as characteristic parameters or coupling constants; it is observed that when $\beta \to \infty$, the influence of the electromagnetic field is negligible. Conversely, in the extremal case where $\beta = 0$, it simplifies to the EMD coupling described by $K(\phi) = e^{2\alpha\phi}$.
The equations of motion are obtained by varying the action with respect to the metric, dilaton, and Maxwell field, respectively, as follows
\be
\1\{\begin{split}
	\label{2}
    &0=\nabla_{\mu}\left[K(\phi)F^{\mu\nu}\right]\\
	&\Box\phi=\frac{1}{4}\frac{\partial K(\phi)}{\partial \phi}F^{2}\\
	&R_{\mu\nu}=2\partial_{\mu}\phi\partial_{\nu}\phi+2K(\phi)
	\left(F_{\mu\sig}F^{~\sig}_{\nu}-\frac{1}{4}g_{\mu\nu}F^{2}\right).
\end{split}\2.
\ee

Due to the field equations mentioned above, reference \cite{9} provides a black hole solution, which can be described as
\be\label{3}
\1\{\begin{split}
&\dif s^{2}=-f(r)\dif t^{2}+f^{-1}(r)\dif r^{2}+C(r) (\dif \theta^2+\sin^2\theta \dif \phi^2)\\
	&f(r)=\left(1-\frac{b_{1}}{r}\right)\left(1-\frac{b_{2}}{r}\right)^
	{\frac{1-\al^{2}}{1+\al^{2}}}+\frac{\beta
		Q^{2}}{C(r)}\\
	&C(r)=r^{2} \left(1-\frac{b_{2}}{r}\right)^{\frac{2\al^{2}}{1+\al^{2}}},
\end{split}\2.
\ee
where $b_{1}$ and $b_{2}$ are functions solely of $\alpha$, yielding
\be
\1\{\begin{split}
	\label{4}
	b_{1}&=(1+\sqrt{1-q^{2}(1-\al^{2})})~M\\
	b_{2}&=\frac{1+\al^{2}}{1-\al^{2}}\left[1-\sqrt{1-q^{2}(1-\al^{2})}\right]~M.
\end{split}\2.
\ee
with $q\equiv{Q}/{M}$ the charge-to-mass ratio, and $M$ the mass of the black hole.

The location of horizons is defined where $f(r_{\pm}) = 0$. Specifically, for
$\beta = 0$, these configurations align with the GMGHS solutions, as referenced
in \cite{11}, where $r_{\pm}$ are identified as $b_1$ and $b_2$, respectively.
Additionally, when $\alpha = 0$, the solution represent the horizons of a
Reissner-Nordstrom(RN) black hole. Conversely, setting $q = 0$ simplifies the
solution of the Schwarzschild black hole. Notably, in scenarios the
EMS model, the horizons emerge strictly under the conditions $0
< q < \sqrt{2}$ and $0 < \beta < \frac{(2-q^2)^2}{4q^2}$ \cite{10,45}.

\section{Geodesic equations and innermost stable circular orbit }
In the present section, we will briefly review the basic results concerning the particle moving on the structure of the metric for a stationary axisymmetric spacetime.  Let's examine an arbitrary symmetric geometry, described by the line element
\be
\label{6}
\begin{split}
\dif s^{2}&=g_{tt}\dif t^2+2g_{t\phi}\dif t\dif \phi+g_{rr}\dif r^2+g_{\th\th}\dif \theta^2+g_{\phi\phi}\dif \phi^2.
\end{split}
\ee
The metric \eqref{6} is suited for spherically symmetric spacetime featuring time-like and space-like Killing vectors $(\partial/\partial t)^\mu$ and $(\partial/\partial \phi)^\mu$ governing time translations and spatial rotations, respectively. In the equatorial approximation $(|\theta-\pi/2|\ll 1)$, the metric functions $g_{tt}$, $g_{t\phi}$, $g_{rr}$, $g_{\th\th}$, and $g_{\phi\phi}$ depend solely on the variable $r$. Within this geometry, it's apparent that motion conserves two quantities: the specific energy at infinity denoted as $\tilde{E}$, and the $z$-component of the specific angular momentum $\tilde{L}$ at infinity \cite{12}. These quantities can be obtained by
\be
\1\{\begin{split}
\label{7}
&g_{tt}\frac{\dif t}{\dif \tau}+g_{t\phi}\frac{\dif \phi}{\dif \tau}=-\tilde{E}\\
&g_{t\phi}\frac{\dif t}{\dif \tau}+g_{\phi\phi}\frac{\dif \phi}{\dif \tau}=\tilde{L},
\end{split}\2.
\ee
where $\tau$ denotes the affine parameter. In the equatorial plane characterized by $\th = \frac{\pi}{2}$, the geodesic equation can be derived as follows
\be
\1\{\begin{split}
\label{8}
&\frac{\dif t}{\dif \tau}=\frac{\tilde{E} g_{\phi\phi}+\tilde{L} g_{t\phi}}{g^2_{t\phi}-g_{tt}g_{\phi\phi}}\\
&\frac{\dif \phi}{\dif \tau}=-\frac{\tilde{E} g_{t\phi}+\tilde{L} g_{tt}}{g^2_{t\phi}-g_{tt}g_{\phi\phi}}\\
&g_{rr}\(\frac{\dif r}{\dif \tau}\)^2=V_{eff}(r),
\end{split}\2.
\ee
with the effective potential
\be
\label{9}
V_{eff}=\frac{\tilde{E}^2g_{\phi\phi}+2\tilde{E}\tilde{ L} g_{t\phi}+\tilde{L}^2 g_{tt}}{g^2_{t\phi}-g_{tt}g_{\phi\phi}}-1.
\ee

The effective potential can be analogized to the Newtonian gravitational
potential, as it provides the condition under which particles undergo circular
orbital motion. The circular orbit within the equatorial plane
($\th=\frac{\pi}{2}$) is defined by the conditions $V_{eff}(r) = 0$ and
$V_{eff,r}(r) = 0$. These conditions establish the specific energy $\tilde{E}$
and the specific angular momentum $\tilde{L}$ as functions of the angular
velocity $\Omega$ of particles (see Appendix B for specific expressions)

\begin{equation}
\left\{\begin{array}{l}
\tilde{E}=\frac{-g_{t t}-\Omega g_{t \phi}}{\sqrt{-g_{t t}-2 \Omega g_{t
\phi}-\Omega^2 g_{\phi \phi}}} \\
\tilde{L}=\frac{g_{t \phi}+\Omega g_{\phi \phi}}{\sqrt{-g_{t t}-2 \Omega g_{t
\phi}-\Omega^2 g_{\phi \phi}}},
\end{array}\right.
\end{equation}
where the angular velocity of the test particle $\Omega$ is
\be
\label{11}
\Omega=\frac{\dif \phi}{\dif t}=\frac{  -g_{t\phi,r} \pm \sqrt{(-g_{t\phi,r})^2-g_{\phi\phi,r} g_{tt,r} }       }{g_{\phi\phi,r}} .
\ee
Furthermore, to ascertain the inner edge of the disk, it is imperative to identify the innermost stable circular orbit (ISCO) of the black hole \cite{21}.
 This involves utilizing the condition $V_{eff,rr}(r) = 0$, which leads to the derivation of the following relation.
\be
\label{12}
\tilde{E}^2 g_{\phi\phi,rr}+2\tilde{E}\tilde{ L} g_{t\phi,rr}+\tilde{L}^2 g_{tt,rr}-(g^2_{t\phi}-g_{tt}g_{\phi\phi})_{,rr}=0 .
\ee
For $V_{eff,rr}(r) <0$, the equatorial circular orbits are stable, so $r_{isco}$ determines the inner edge of the thin accretion disk \cite{22}. In Fig.1 and Fig.2, we depict the change in particle angular velocity, specific energy, specific angular momentum, and $r_{isco}$ concerning the distance $r$ and the black hole EMS parameters $\alpha$, $\beta$.

 According to Fig.1, it is evident that the angular velocity of the EMS
	black
	hole exceeds that of the RN black hole for a fixed distance $r$,
	charge-to-mass
	ratio $q=0.5$, and parameter $\beta=0$. Furthermore, as the distance $r$
	increases, the angular velocity shows a declining pattern. However, the
	specific energy $\tilde{E}$ and specific angular momentum $\tilde{L}$
	demonstrate distinct patterns. It is noteworthy that both the specific
	energy
	$\tilde{E}$ and specific angular momentum $\tilde{L}$ exhibit a decreasing
	trend followed by an increasing trend, respectively. Furthermore, from the
	last
	diagram, we obtain that within the allowed parameter $\alpha$ and $\beta$,
	as
	the parameter $\beta$ increases, the innermost stable circular orbit
	gradually
	decreases for different $\alpha$.
\begin{figure}[H]
	\begin{minipage}{0.5\textwidth}
		\includegraphics[scale=0.83,angle=0]{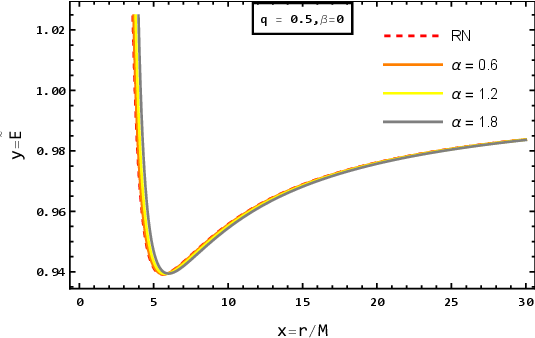}
	\end{minipage}%
	\begin{minipage}{0.5\textwidth}
		\includegraphics[scale=0.8,angle=0]{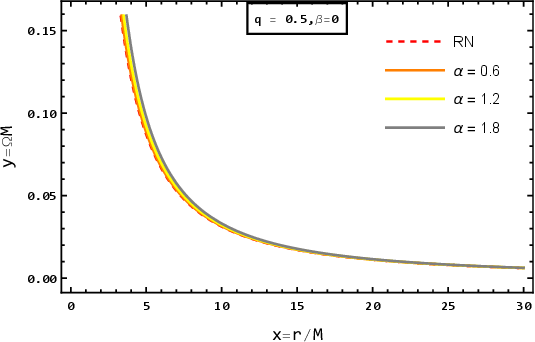}
	\end{minipage}	
	\begin{minipage}{0.5\textwidth}
		\includegraphics[scale=0.8,angle=0]{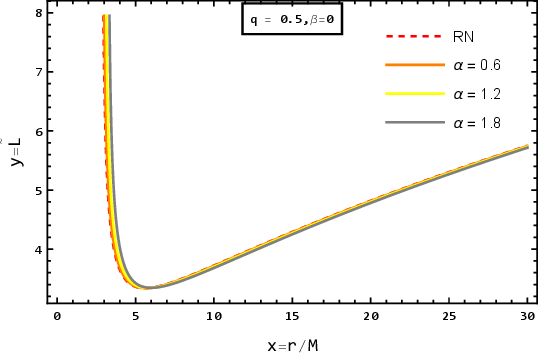}
	\end{minipage}%
	\begin{minipage}{0.5\textwidth}
		\includegraphics[scale=0.8,angle=0]{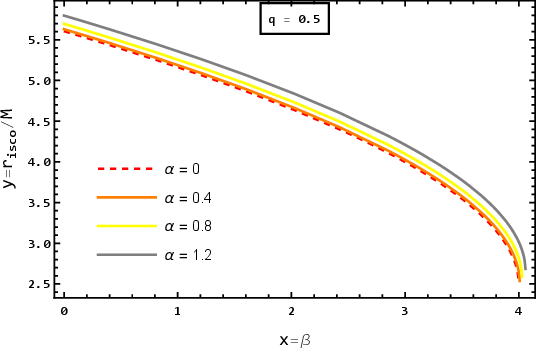}
	\end{minipage}
	\caption{\label{Fig.1} The above figures respectively depict the variation
	curves of energy, angular velocity, and angular momentum with different
	values of $\alpha$.  The bottom-right panel illustrates the variation of
	$r_{ISCO}$ under different EMS parameters $\alpha$.}
 \end{figure}

Similar to the trends observed in Fig.1, the findings from Fig.2 reveal
that, when $q=0.5$ and $\alpha=0$ are held constant, both the specific energy
$\tilde{E}$ and specific angular momentum $\tilde{L}$ initially decrease with
increasing distance $r$ before exhibiting an increasing trend. Meanwhile, the
angular velocity decreases as $r$ increases. When $r$ is kept fixed, it becomes
evident that the RN black hole demonstrates higher values for specific angular
momentum, specific energy, and angular velocity compared to the EMS black hole.
However, contrary to the observations in Fig. 1, the plot depicting $r_{isco}$
shows an increasing trend with higher values of $\alpha$, while $q=0.5$ remains fixed.

In addition, we investigated the effects of RN black hole
compared to Schwarzschild black hole. As shown in Fig.3, with an
increasing charge-to-mass ratio $q$ of the RN black hole, the specific energy,
specific angular momentum, angular velocity, and the second derivative of the
potential function all decrease.

\begin{figure}[H]
	\begin{minipage}{0.45\textwidth}
		\includegraphics[scale=0.8,angle=0]{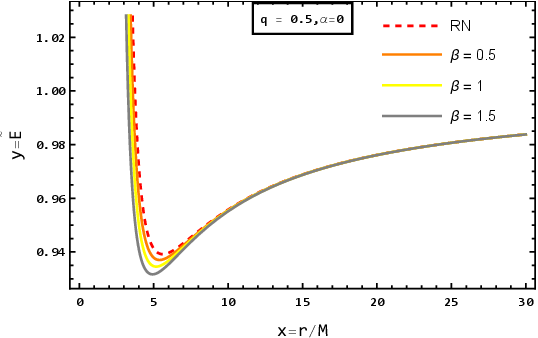}
	\end{minipage}%
	\begin{minipage}{0.9\textwidth}
		\includegraphics[scale=0.8,angle=0]{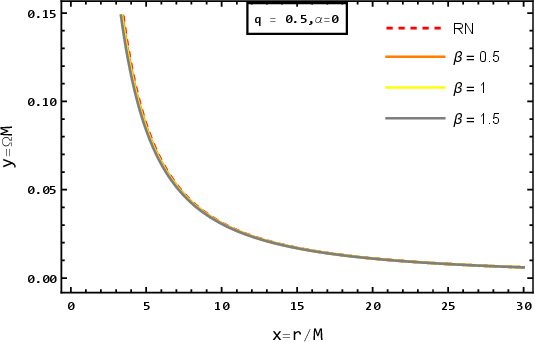}
	\end{minipage}	
	\begin{minipage}{0.45\textwidth}
		\includegraphics[scale=0.8,angle=0]{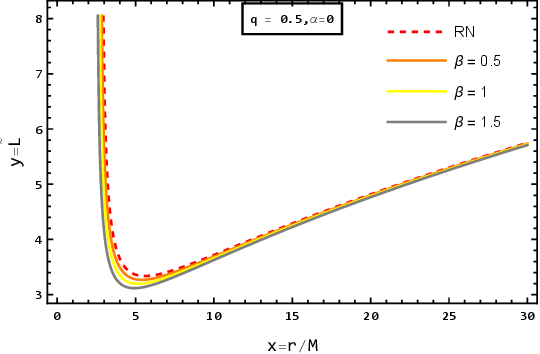}
	\end{minipage}%
	\begin{minipage}{0.9\textwidth}
		\includegraphics[scale=0.8,angle=0]{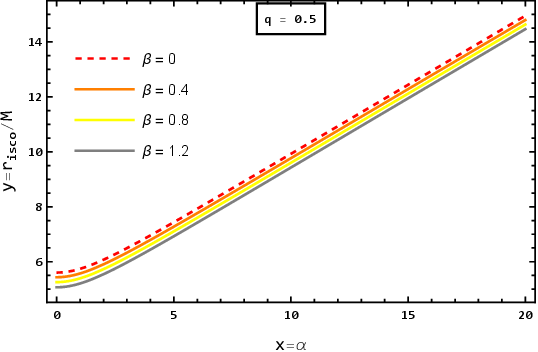}
	\end{minipage}
	\caption{\label{Fig.2} The above four plots also depict the
	variation curves of energy, angular velocity, angular momentum, and ISCO
	with different parameters $\beta$.}
\end{figure}

\begin{figure}[H]
	\begin{minipage}{0.45\textwidth}
		\includegraphics[scale=0.8,angle=0]{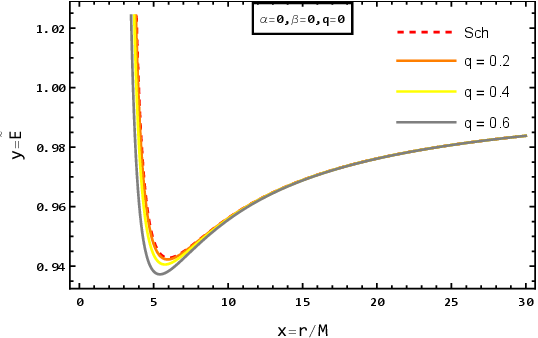}
	\end{minipage}	
	\begin{minipage}{0.9\textwidth}
		\includegraphics[scale=0.8,angle=0]{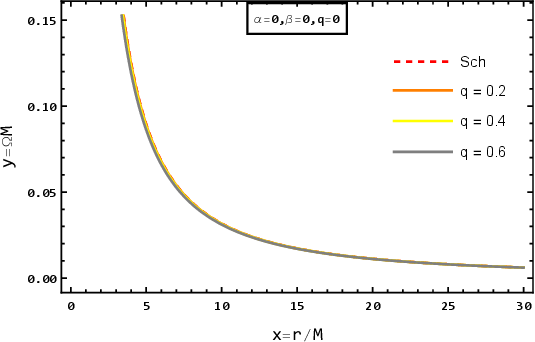}
	\end{minipage}%
	
	\begin{minipage}{0.45\textwidth}
		\includegraphics[scale=0.8,angle=0]{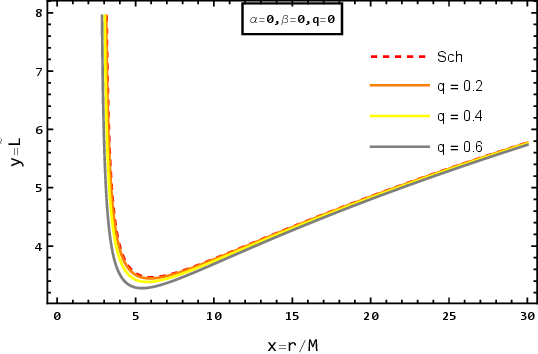}
	\end{minipage}
	\begin{minipage}{0.9\textwidth}
		\includegraphics[scale=0.8,angle=0]{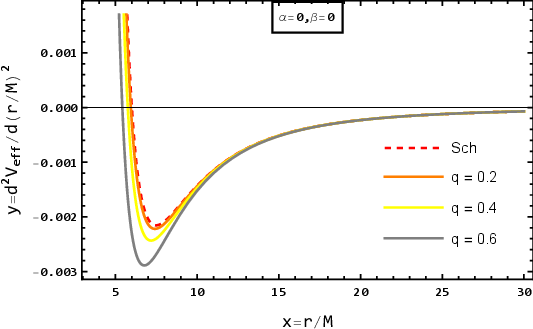}
	\end{minipage}%
	\caption{\label{Fig 2}The figures depict a comparison between RN and Schwarzschild black holes. Parameters $\alpha$ and $\beta$ are set to $\alpha=0$ and $\beta=0$. }
\end{figure}

\section{The properties of thin accretion disks in Einstein-Maxwell-scalar theory}
The thin accretion disk surrounding a black hole is a structure formed around
the gravitational singularity, where celestial matter rotates in a thin
disk-like configuration. These accretion disks represent one of the most active
and dynamic regions around black holes, influenced by complex factors such as
gravity, magnetic fields, and material viscosity. Typically consisting of
discrete rings, these disks rotate rapidly under the influence of the black
hole's gravity, emitting intense radiation, including X-rays and gamma-rays.
The temperature gradient within the accretion disk increases with distance from
the center of the black hole, leading to complex spectral features in the
emitted radiation. Studying the properties of thin accretion disks around black
holes is of paramount importance for understanding the accretion processes, jet
formation mechanisms, and the formation of the most extreme astrophysical
phenomena in the universe.

In this section, we begin by offering a concise overview of the physical
characteristics of thin accretion disks, as outlined by the Novikov-Thorne
model \cite{19}, which builds upon the Shakura-Sunyaev model \cite{20}. This
model is built upon several fundamental assumptions: (1) the spacetime
surrounding the central massive object is stationary, axisymmetric, and
asymptotically flat; (2) the mass of the disk does not perturb the background
metric; (3) the accretion disk is geometrically thin, with its vertical
dimension considered negligible compared to its horizontal extent; (4) orbiting
particles around the central compact object travel between the outer edge
$r_{\text{out}}$ and $r_{\text{isco}}$, establishing the inner boundary of the
disk; (5) the accretion disk lies in the equatorial plane of the accreting
compact object, perpendicular to the black hole's spin axis; (6) due to
hydrodynamic and thermodynamic equilibrium within the disk, the electromagnetic
radiation emitted from its surface is assumed to exhibit a black body spectrum;
(7) the mass accretion rate of the disk, $\dot{M}_0$, remains constant over
time.

Subsequently, we undertake a thorough investigation into the radiant energy flux, radiation temperature, observed luminosity, and other relevant parameters associated with the thin accretion disk. Despite encountering numerical complexities in solving the equation $V_{\mathrm{eff}, \mathrm{r}}=0$, and the necessity for the spacetime described in the Novikov-Thorne model to approach asymptotic flatness, we concentrate on the scenario with zero cosmological constant for simplicity, ensuring generality without sacrificing accuracy. We employ the following values for physical constants and the thin accretion disk: $c=2.997 \times 10^{10} \mathrm{cms}^{-1}, \dot{M}_0=$ $2 \times 10^{-6}M_{\odot \mathrm{yr}^{-1}}, 1 \mathrm{yr}=3.156 \times 10^7 \mathrm{~s},\sigma_{\mathrm{SB}}=5.67 \times 10^{-5} \mathrm{erg} \mathrm{s}^{-1}\mathrm{~cm}^{-2} \mathrm{~K}^{-4}, h=6.625 \times 10^{-27} \mathrm{ergs},k_{\mathrm{B}}=1.38 \times 10^{-16}$ $\operatorname{ergK}^{-1}, M_{\odot}=1.989\times 10^{33} \mathrm{~g}$, and the mass $M=2 \times 10^6M_{\odot}$.
\subsection{The radiant energy flux}
From the conservation equations of rest mass, energy, and the angular momentum of the disk particles, the radiant energy flux over the disk surface can be obtained as \cite{13,14}
\begin{equation}
F(r)=-\frac{\dot{M}}{4 \pi \sqrt{-\tilde{g}}} \frac{\Omega_{, r}}{(\tilde{E}-\Omega \tilde{L})^2} \int_{r_{isco}}^r(\tilde{E}-\Omega \tilde{L}) \tilde{L}_{, r} \mathrm{~d} r .
\end{equation}
This equation was widely used in the literature. It is valid only for cylindrical coordinates. If adopting spherical coordinates, it takes
the form (see Appendix C for derivations)
\begin{equation}
F(r)=-\frac{c^2 \dot{M}}{4 \pi \sqrt{-g / g_{\theta \theta}}} \frac{\Omega_{, r}}{(\tilde{E}-\Omega \tilde{L})^2} \int_{r_{isco}}^r(\tilde{E}-\Omega \tilde{L}) \tilde{L}_{, r} \mathrm{~d} r,
\end{equation}
Here, we've recovered dimensions, where $c$ represents the speed of light. We investigate mass accretion driven by black holes with a total mass of $M=2\times10^6M_{\odot}$ and a corresponding mass accretion rate of $\dot{M}=2\times10^{-6}M_{\odot}yr^{-1}$.

Fig.4 presents the energy flux $F(r)$ emitted from the disk encircling an EMS black hole with different parameters $\alpha$ and $\beta$. Remarkably, the energy flux distribution displays a distinctive pattern of initial ascent, peak attainment, and subsequent descent. The horizontal axis delineates the dimensionless coordinate distance $r$, while the vertical axis describes the energy flux magnitude. Furthermore, we observe that, holding constant parameters $q=0.5$ and $\beta=0$, as well as the distance $r$, the radiation flux decreases with the increase of parameter $\alpha$. However, when keeping parameters $q=0.5$ and $\alpha=0$ fixed, an increase in parameter $\beta$ results in a gradual increase in radiation flux $F(r)$. The third figure provides a comparison between RN black holes and Schwarzschild black holes. It is clear that charge has a direct impact on the physical quantities, with an increasing charge-to-mass ratio $q$ resulting in a rise in radiation flux.
\begin{figure}[H]
	\begin{minipage}{0.5\textwidth}
		\includegraphics[scale=0.9,angle=0]{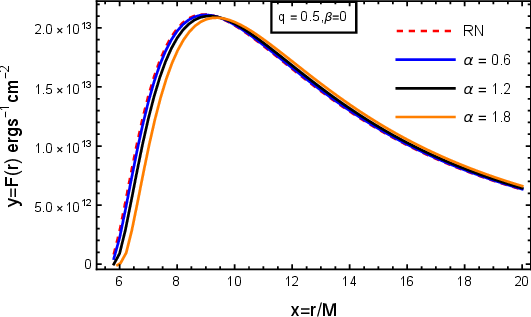}
	\end{minipage}%
	\begin{minipage}{0.5\textwidth}
		\includegraphics[scale=0.9,angle=0]{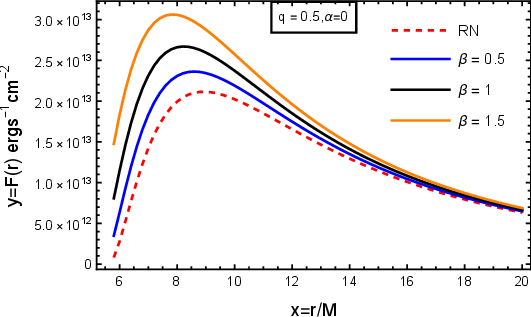}
	\end{minipage}

\begin{minipage}{0.45\textwidth}
	\includegraphics[scale=0.9,angle=0]{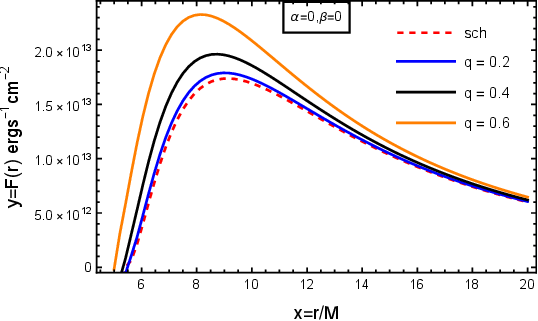}
\end{minipage}		
	\caption{\label{Fig.3}The energy flux $F(r)$ around the EMS black hole for different values of  parameters $\alpha$ and $\beta$. we set the mass of sun $M_{\odot}= 1.989 \times10^{33}$g. And the energy flux, the bottom-left panel, is plotted against values of different $q$. The dashed line represents the radiation scenario for the RN and Schwarzschild black holes. }
\end{figure}
\subsection{The radiation temperature}
In the framework of the Novikov-Thorne model, the accreted material attains a state of thermodynamic equilibrium, implying that the radiation emitted by the disk behaves similarly to perfect black body radiation. The radiation temperature $T(r)$ of the disk is connected to the energy flux $F(r)$ via the Stefan-Boltzmann law, expressed as $F(r)=\sigma_{\mathrm{SB}} T^4(r)$, where $\sigma_{\mathrm{SB}}$ denotes the Stefan-Boltzmann constant. This suggests that the variation of $T(r)$ with respect to $r$ corresponds to the dependence of the energy flux $F(r)$ on $r$. The radiation temperature $T(r)$ surrounding the EMS black hole has been plotted in Fig.5 . Similarly, we systematically fix one parameter within the EMS theory while investigating the influence of the other parameter on the radiation temperature $T(r)$. From the left plot (with $q=0.5$,$\beta=0$), it is evident that as the parameter $\al$ increases, $T(r)$ decreases. Similarly, from the right plot (with $q=0.5$,$\al=0$), it can be inferred that as the parameter $\beta$ increases, $T(r)$ increases. Additionally, the figure below shows that an increase in the charge-to-mass ratio $q$ of the RN black hole enhances the radiation temperature. The overall temperature plot exhibits a trend of initially increasing and then decreasing. Conversely, for parameter $\beta=0$ and $q=0.5$, the disk is cooler compared to the cases of RN black hole.
\begin{figure}[H]
	\begin{minipage}{0.5\textwidth}
		\includegraphics[scale=0.9,angle=0]{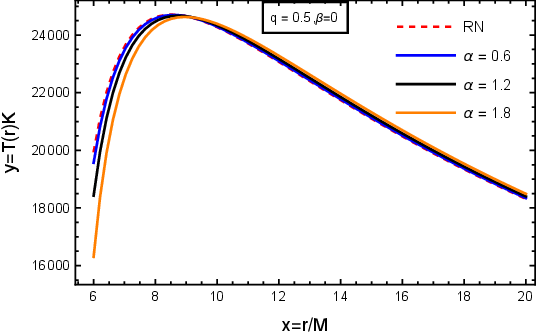}
	\end{minipage}%
	\begin{minipage}{0.5\textwidth}
		\includegraphics[scale=0.9,angle=0]{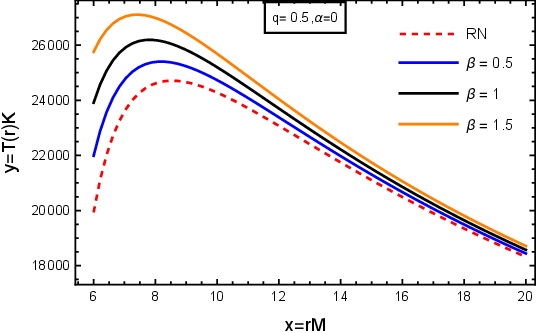}
	\end{minipage}	

\begin{minipage}{0.43\textwidth} 
	\includegraphics[scale=0.9,angle=0]{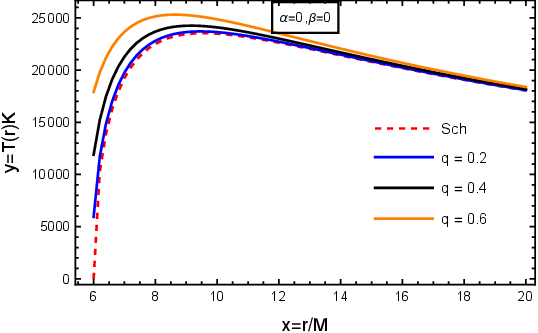}
\end{minipage}
	\caption{\label{Fig.4}
The top two figures depict the impact of parameters $\alpha$ and $\beta$  on the temperature of the thin accretion disk.
 Here, we use the Stefan-Boltzmann constant
	$\sigma_{SB}=5.67 \times 10^{-5} \mathrm{erg}
	\mathrm{s}^{-1}\mathrm{~cm}^{-2} \mathrm{~K}^{-4}$.  And the temperature, the bottom-left panel, is plotted against values of different $q$.  }
\end{figure}
  \subsection{The observed luminosity}
The observed luminosity $L(\nu)$ for the thin accretion disk around a black hole exhibits a redshifted black body spectrum as described in \cite{16}.
\begin{equation}
L(\nu)=4 \pi \mathrm{d}^2 I(\nu)=8 \pi h \cos \gamma
\int_{r_{\mathrm{i}}}^{r_{\mathrm{f}}} \int_0^{2 \pi} \frac{\nu_e^3 r d r d
\varphi}{e^{\frac{h \nu_e}{k_{\mathrm{B}} T}}-1}=8 \pi h \kappa^{\frac{2}{1+2
\mu}} \cos \gamma \int_{\bar{r}_{\mathrm{i}}}^{\bar{r}_{\mathrm{f}}} \int_0^{2
\pi} \frac{\nu_e^3 \bar{r} d \bar{r} d \varphi}{e^{\frac{h
\nu_e}{k_{\mathrm{B}} T}}-1},
\end{equation}
Here, $d$ denotes the distance to the disk center, $I(\nu)$ represents the thermal energy flux radiated by the disk, $h$ stands for Planck's constant, $k_{\mathrm{B}}$ denotes the Boltzmann constant, and $\gamma$ signifies the disk inclination angle, which we set to zero for simplicity. The parameters $r_{\mathrm{f}}$ and $r_{\mathrm{i}}$ refer to the outer and inner radii of the disk edge, respectively. Assuming the flux over the disk surface vanishes as $r_{\mathrm{f}} \rightarrow \infty$, we select $r_{\mathrm{i}}=r_{\text {isco }}$ and $r_{\mathrm{f}} \rightarrow\infty$ to compute the disk's luminosity $L(\nu)$. The emitted frequency is determined by $\nu_e=\nu(1+z)$, where the redshift factor can be expressed as
\begin{equation}
1+z=\frac{1+\Omega r \sin \varphi \sin \gamma}{\sqrt{-g_{t t}-2 \Omega g_{t
\varphi}-\Omega^2 g_{\varphi \varphi}}},
\end{equation}
where the light bending is neglected \cite{17}. As depicted in Fig.6 , the spectral energy distribution of accretion disks around EMS black holes is clearly illustrated.
\begin{figure}[H]
	\begin{minipage}{0.5\textwidth}
		\includegraphics[scale=0.9,angle=0]{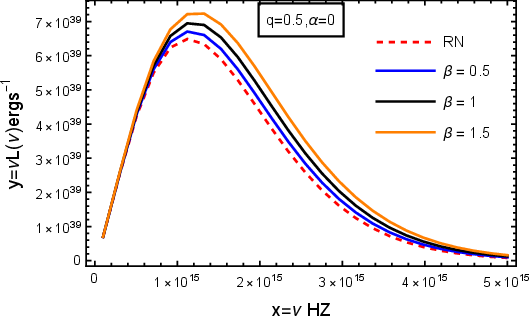}
	\end{minipage}%
	\begin{minipage}{0.5\textwidth}
		\includegraphics[scale=0.9,angle=0]{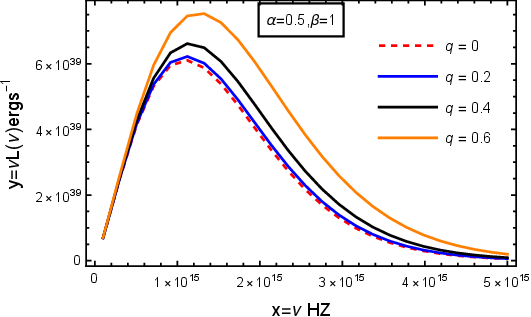}
	\end{minipage}	

\begin{minipage}{0.45\textwidth}
	\includegraphics[scale=0.9,angle=0]{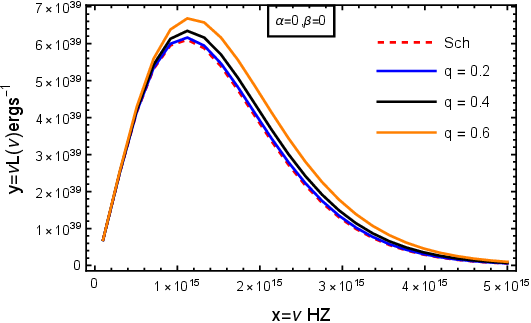}
\end{minipage}	
	\caption{\label{Fig.5}The emission spectrum $\nu L(\nu)$ of disk around EMS black hole with different values of $\alpha$ and $\beta$.  And the emission spectrum, the bottom-left panel, is plotted against values of different $q$. }
\end{figure}

 Similar to the trends observed in Fig.4 and Fig.5 regarding energy flux and radiation temperature, respectively, the spectral energy of the disk around an EMS black hole exhibits higher luminosity compared to that around an RN black hole with fixed parameters $q=0.5$ and $\alpha=0$. Moreover, when $\beta$ is fixed at $0$, there appears to be no discernible difference in luminosity for varying values of $\alpha$. Notably, from the right plot, it is evident that with a fixed parameter setting of $\alpha=0$ and $\beta=1$, an increase in $q$ leads to a corresponding increase in radiation luminosity, surpassing that of the RN black hole. However, as shown in the third figure, the luminosity of the RN black hole exceeds that of the Schwarzschild black hole.

\subsection{The radiative efficiency}

The radiative efficiency of a black hole refers to the ratio of the energy
radiated away by the accretion process to the rest-mass energy of the accreted
material. This efficiency plays a crucial role in determining the luminosity
and spectral characteristics of black hole accretion disks. It is often
quantified by the Eddington luminosity, which represents the maximum luminosity
a black hole can achieve when the radiation pressure balances gravitational
attraction. Understanding and characterizing the radiative efficiency of black
holes are essential in elucidating their accretion processes, as well as their
influence on the surrounding environment and observational signatures across
different wavelengths.  Assuming all emitted photons can reach infinity, the
efficiency $\epsilon$ is governed by the specific energy of a particle at the
marginally stable orbit $r_{isco}$, which follows \cite{18}
\begin{equation}
\epsilon=1-E_{\text {isco }} .
\end{equation}
The Fig.7 and Tab.1 depict the trend of the other model parameter $\alpha$ (or
$\beta$) as the radiative efficiency varies, with $q=0.5$ fixed. Fig.7
indicates that when $\beta$ is fixed, $\alpha$ has almost no effect on the
radiative efficiency $\epsilon$, while when $\alpha$ is fixed, $\beta$ has a
nearly linear impact on the radiative efficiency. In Tab.1, the
first row, where $ q = 0$, corresponds to the Schwarzschild black hole, while
increasing $q$ represents the RN black hole. The results indicate that as $q$
increases, the radiation efficiency of the RN black hole surpasses that of the
Schwarzschild black hole. Furthermore, it is noted that the radiative
efficiency of the EMS black hole exceeds that of the RN black hole with fixed
parameters. This observation enriches our comprehension of these two black holes through empirical observations, shedding light on their distinct
characteristics and behaviors within accretion systems.\\
\begin{table}[H]
	\centering
	\caption{\label{Tab.1} The Table illustrates the variation of
	the radiative efficiency $\epsilon$ derived from the accretion disc.   }
	\setlength{\tabcolsep}{4mm}{
		\begin{tabular}{ccccccccc}
			\hline
			\hline
			$\alpha$  & & $\beta$   &  & $ q $ &  & $r_{isco}/M$    &&
			$\epsilon$             \\
			
			\hline
			0&                             &0
			&& 0 &               &6.000&                 &0.05719             \\
			&                              &
			&& 0.2 &              &5.940&
			&0.05772                   \\
			&                             &                                  &&
			0.4 &              &5.753&                &0.05943                \\
			&                              &
			&& 0.6 &              &5.420&
			&0.06272                   \\
			\hline
			0 &
			&0                                   && 0.5 &         &5.607
			&               & 0.06083                         \\
			0.6&
			&                                   &&  &               &5.658
			&                &0.06080                 \\
			1.2&
			&                                   &&  &               &5.797
			&                &0.06071                   \\
			1.8&
			&                                   &&  &               &5.997
			&                &0.06058             \\
			\hline
			0 &
			&0                                   && 0.5 &            &5.607
			&               & 0.06083                       \\
			&
			&0.5                                   &&  &               &5.393
			&                &0.06299                 \\
			&                             &1
			&&  &                &5.166   &
			&0.06547                 \\
			&                             &
			1.5                                  &&  &              &4.921
			&                &0.06834            \\
			\hline
	\end{tabular}}
\end{table}

\begin{figure}[H]
	\centering
	\begin{minipage}{0.5\textwidth}
		\centering
		\includegraphics[scale=0.9,angle=0]{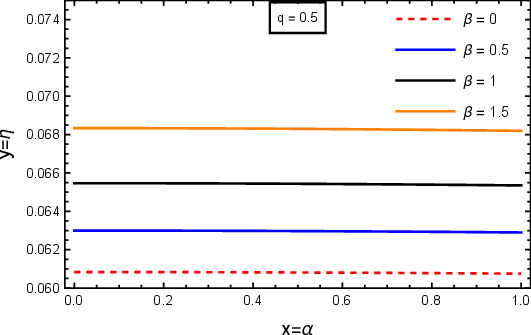}
	\end{minipage}%
	\begin{minipage}{0.5\textwidth}
		\centering
		\includegraphics[scale=0.9,angle=0]{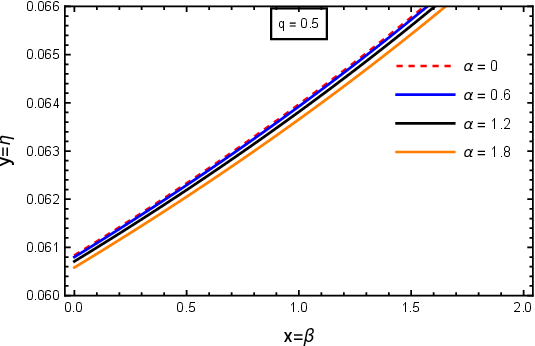}
	\end{minipage}	
	\caption{\label{Fig.6}
The two figures above depict the impact of parameters $\alpha$ and $ \beta$ on the radiative efficiency  of the thin accretion disk.}
\end{figure}

\section{Conclusion and discussion}
In our study of spherically symmetric black holes within the
Einstein-Maxwell-scalar(EMS) framework, we examined the dynamics of thin
relativistic accretion disks utilizing the Novikov-Thorne model. We began by
deriving essential physical quantities such as the effective potential
$V_{\text{eff}}$, specific angular momentum $\tilde{L}$, specific energy
$\tilde{E}$, angular velocity $\Omega$, and the radius of the innermost stable
circular orbit (ISCO) for test particles in circular orbits around the black
hole. Due to the complexity of obtaining an analytical expression for the ISCO radius, we employed numerical methods to calculate it for various EMS
parameters $\alpha$ and $\beta$.

Our findings indicate that, holding the charge-to-mass ratio $q$ constant, the ISCO radius extends as $\al$ increases, and contracts as $\beta$ rises. Additionally, the detected particles $\tilde{L}$ and $\tilde{E}$ transform with changes in $\alpha$ and $\beta$. The graphical representations of specific energy and specific angular momentum display a pattern where both initially decrease to a minimum before ascending, while the angular velocity predominantly diminishes. Notably, when $q=0.5$, $\beta=0$, the values of $\tilde{E}$, $\tilde{L}$, and $\Omega$ exceed those of the RN black hole. Conversely, when  $q=0.5$, $\al=0$, these physical quantities are lower than those associated with the RN black hole.  When the parameters $\alpha$ and $\beta$ are both fixed at 0, an increase in the charge-to-mass ratio $q$ of the RN black hole leads to a decrease in the specific energy, specific angular momentum, and angular velocity. This highlights significant differences in the dynamic properties of these black holes under varying parameter conditions.

Additionally, we performed numerical calculations to determine the heat emitted by the accretion disk, obtaining the energy flux $F(r)$, temperature $T(r)$, radiative efficiency $\epsilon$, and luminosity distribution $L(\nu)$. Diagrams for these quantities were subsequently created, illustrating the influence of the EMS parameters on these observable quantities. Under fixed EMS parameters $\beta$ and $q$, we observed that with increasing $\alpha$, both the radiation flux and temperature decrease and are lower than the results for the RN black hole. However, the luminosity and radiative efficiency remain nearly unchanged. Similarly, when $\alpha$ and $q$ are fixed, we found that the radiation flux, temperature, luminosity, and radiative efficiency increase with increasing $\beta$.  In the same way, with the parameters fixed at $\alpha=\beta=0$, the radiation flux, temperature, and luminosity of the RN black hole surpass those of the Schwarzschild black hole. It can be concluded that the model parameters have varying effects on the results, which will also contribute to observations of different black holes.

\begin{acknowledgments}
This work was supported by the National Key R\&D Program of China (Grants No. 2022YFA1403700),  NSFC (Grants No. 12141402, 12334002), the SUSTech-NUS Joint Research Program, and Center for Computational Science and Engineering at Southern University of Science and Technology.
\end{acknowledgments}

\appendix
\section{Derivations for $K(\phi)$}
In this part, we derive the Eq.(2) following the paper \cite{100}. The
metric
for static and spherically symmetric black hole solutions can always be written
as equation
\begin{equation}
d s^2=-f(r) d t^2+f(r)^{-1} d r^2+C(r)\left(d \theta^2+\sin ^2 \theta d
\phi^2\right).
\end{equation}
In this spacetime, the non-vanishing components of four-vector
$A_{\mu}$ is uniquely $A_{0}(r)$. Furthermore, the equations of motion can turn
out to be
\begin{equation}
\left\{
\begin{aligned}
&2 CC^{\prime \prime}+4 C^2 \phi^{\prime 2}-C^{\prime 2}=0, \\
&\left(C K A_0^{\prime}\right)^{\prime}=0, \\
&2 f^2 f^{\prime \prime}+2 C f C^{\prime \prime}+2 C f^{\prime} C^{\prime}-f
C^{\prime 2}+4 f C^2 \phi^{\prime 2} \\
&-4 C^2 K A_0^{\prime 2}+2 C^2 V=0, \\
&C f \phi^{\prime \prime}+f C^{\prime} \phi^{\prime}+C f^{\prime}
\phi^{\prime}+\frac{1}{2} C K_{, \phi} A_0^{\prime 2}-\frac{1}{4} C V_{,
\phi}=0.
\end{aligned} \label{300}
\right.
\end{equation}
Here prime denotes the derivative concerning $r$.  We remember that the
dilaton black hole in de-Sitter universe for arbitrary
coupling constant
$\alpha$ is
\begin{equation}
\left\{
\begin{aligned}
&f=\left(1-\frac{b_{1}}{r}\right)\left(1-\frac{b_{2}}{r}\right)^{\frac{1-\alpha^2}{1+\alpha^2}}-\frac{1}{3}
\lambda C,\\
&C=r^2\left(1-\frac{b_{2}}{r}\right)^{\frac{2 \alpha^2}{1+\alpha^2}}.
\end{aligned}\label{100}
\right.
\end{equation}
Here $b_{1}$ and $b_{2}$ are two constants that are determined by the black hole mass
$M$,
charge $Q$ and
coupling constant $\alpha$. The corresponding coupling function $K$ and scalar
potential $V$ in the action are
\begin{equation}
\left\{
\begin{aligned}
&K=e^{2 \alpha \phi},\\
&V=\frac{2 \lambda}{3\left(1+\alpha^2\right)^2}\left[\alpha^2\left(3
\alpha^2-1\right) e^{2 \phi / \alpha}+\left(3-\alpha^2\right) e^{-2 \alpha
	\phi}\right. \\
& \left.+8 \alpha^2 e^{-\phi \alpha+\phi / \alpha}\right].
\end{aligned}
\right. \label{111}
\end{equation}
Observing the $\lambda$ term in Eq. (\ref{100}),  we find it is proportional to
$C$. So we presume
\begin{equation}
\left\{
\begin{aligned}
&f=\left(1-\frac{b_{1}}{r}\right)\left(1-\frac{b_{2}}{r}\right)^{\frac{1-\alpha^2}{1+\alpha^2}}+\frac{\beta
	Q^2}{C}-\frac{1}{3} \lambda C ,\\
&C=r^2\left(1-\frac{b_{2}}{r}\right)^{\frac{2 \alpha^2}{1+\alpha^2}}.
\end{aligned} \label{200}
\right.
\end{equation}
Substituting Eq.(\ref{200}) into the equations of motion (\ref{300}), we obtain
\begin{equation}
\left\{
\begin{aligned}
&K(\phi)=\frac{e^{2 \alpha
\phi}\left(\alpha^2+1\right)}{\alpha^2+\beta+1+\alpha^2
	\beta e^{\frac{2 \phi\left(\alpha^2+1\right)}{\alpha}}}.\\
&V=\frac{2 \lambda}{3\left(1+\alpha^2\right)^2}\left[\alpha^2\left(3
\alpha^2-1\right) e^{2 \phi / \alpha}+\left(3-\alpha^2\right) e^{-2 \alpha
	\phi}\right. \\
& \left.+8 \alpha^2 e^{-\phi \alpha+\phi / \alpha}\right].
\end{aligned} \label{222}
\right.
\end{equation}

Eq.(\ref{222}) can be seen as a generalization of Eq.(\ref{111}), applicable to
generic spherically symmetric black holes with a general coupling $K(\phi)$.

\section{Expressions for some formulas}
In this appendix, we present the explicit analytical forms of the specific
energy, angular momentum, Keplerian angular velocity, and effective potential.
Firstly, we define the following equations,
\begin{equation}
\left\{
	\begin{aligned}
	&f_{1}\equiv \sqrt{\left(\alpha ^2-1\right) q^2+1}-1,\\
	&f_{2}\equiv \left(1-\frac{\left(\alpha ^2+1\right) M
		\left(\sqrt{\left(\alpha ^2-1\right) q^2+1}-1\right)}{\left(\alpha
		^2-1\right) r}\right)^{-\frac{2 \alpha ^2}{\alpha ^2+1}},\\
	&f_{3}\equiv	f_1 M^2 q^2 \left(\alpha ^2+\beta +1\right)+M r
	\left(-\left(\alpha
	^2+1\right) f_1-\left(\left(\alpha ^2-1\right) (\beta +1)
	q^2\right)\right)+\left(\alpha ^2-1\right) r^2,\\
	&f_{4}\equiv r^2-\frac{2 M r \left(\alpha ^2 \sqrt{\left(\alpha ^2-1\right)
			q^2+1}-1\right)}{\alpha ^2-1}.
	\end{aligned}\right.
\end{equation}

Then, we present  the explicit analytical forms of the specific
energy, angular momentum, Keplerian angular velocity, and effective potential.
\begin{equation}
	\left\{
	\begin{aligned}
	&\Omega=\frac{f_2 \left(\left(\alpha ^2-1\right) r-\left(\alpha ^2+1\right)
		f_1
		M\right) \sqrt{-\left(f_3 M \left(f_1 M+\alpha ^2
			(-r)+r\right)\right)}}{r^2 \left(\left(\alpha ^2+1\right) f_1
			M+\alpha
		^2
		(-r)+r\right) \left(M+Mf_1+\left(\alpha ^2-1\right) r\right)},\\
	&\tilde{E}=\frac{f_2 \left(f_4+M^2 q^2 \left(\alpha ^2+\beta
		+1\right)\right)}{r
		\sqrt{f_2 \left(-\frac{r^4 \Omega ^2}{f_2^2}+f_4+M^2 q^2 \left(\alpha
			^2+\beta
			+1\right)\right)}},\\
	&\tilde{L}=\frac{f_2 r^3 \Omega }{\sqrt{f_2 \left(-\frac{r^4 \Omega
				^2}{f_2^2}+f_4+M^2 q^2 \left(\alpha ^2+\beta
				+1\right)\right)}},\\
	&V_{eff}=	\frac{\frac{f_2 \tilde{L}^2 \left(-f_4-M^2 q^2 \left(\alpha
			^2+\beta
			+1\right)\right)}{r^2}+\frac{2 f_1 M r}{\alpha
			^2-1}+\left(\frac{\tilde{E}^2}{f_2}-1\right) r^2-M^2 q^2
			\left(\alpha
		^2+\beta
		+1\right)}{f_4+M^2 q^2 \left(\alpha ^2+\beta +1\right)}.
	\end{aligned}
	\right.
\end{equation}

\section{Radiant Energy Flux in Adapted Spherical Coordinates}
In Appendix C, we just show how we obtain the $\sqrt{-g/g_{\theta\theta}}$. The
equations describing thin accretion disks widely used in
the literature were first derived in Novikov \& Thorne (1973),
Page \& Thorne (1974), where the authors employed adapted
cylindrical coordinates near the equatorial plane. Some of these
equations are revisited here for a metric written over adapted
spherical coordinates because there are a few subtleties one needs to heed
\cite{101}. The
end result, though, is exactly the same in
this approximation regime.

As in the original paper, quantities written inside brackets are
averaged over time and the axial coordinate:
\begin{equation}
\langle\Psi(r, \theta)\rangle \equiv \frac{1}{2 \pi \Delta t} \int_0^{\Delta t}
\int_0^{2 \pi} \Psi(t, r, \theta, \varphi) d \varphi d t .
\end{equation}

The four-velocity of the fluid in its local rest frame, $u_{\text {inst
}}^\mu$, is also averaged over the height of the disk and weighted by its rest
mass density $\rho_0$

\begin{equation}
u^\mu \equiv \frac{1}{\Sigma} \int_{-H}^{H}\left\langle\rho_0
u_{\text {inst }}^\mu\right\rangle dz,
\end{equation}
Here, $H$ denotes the maximum half-thickness of the disk. It is important to
emphasize that this definition is expressed in cylindrical coordinates. To
convert it into spherical coordinates, a Jacobian transformation must be
applied.

\begin{equation}
u^\mu= \frac{1}{\Sigma} \int_{\pi-\theta_0}^{\theta_0}\left\langle\rho_0
u_{\text {inst }}^\mu\right\rangle \sqrt{g_{\theta \theta}} d \theta,
\end{equation}

where $\Sigma$ is the averaged surface energy density at a certain point:

\begin{equation}
\Sigma(r) \equiv
\int_{\pi-\theta_0}^{\theta_0}\left\langle\rho_0\right\rangle
\sqrt{g_{\theta \theta}} d \theta.
\end{equation}

The energy-momentum tensor of the fluid is written in terms of the averaged
four-velocity in general form as
\begin{equation}
T^{\mu \nu}=\rho_0 u^\mu u^\nu+t^{\mu \nu}+2 u^{(\mu} q^{\nu)},
\end{equation}
where $t^{\mu \nu}$ is the averaged energy-momentum tensor in the rest frame
and is orthogonal to $u^\mu$, and $q^\mu$ is an energy flux vector field, also
orthogonal to the four-velocity.

Among all of the assumptions made and used to derive the equations is the
assertion that the disk is optically thick, i.e., the radiation emitted along
the disk's plane is negligible, and one considers only the flux orthogonal to
its face (hence, $q^\mu u_\mu=0$ ). Thus, the average flux of radiant energy
is given by
\begin{equation}
F(r) \equiv\left\langle q^\theta\left(r, \theta_0\right) \sqrt{g_{\theta
	\theta\left(r, \theta_0\right)}}\right\rangle=\left\langle q^\theta\left(r,
\pi-\theta_0\right) \sqrt{g_{\theta \theta\left(r,
	\pi-\theta_0\right)}}\right\rangle,
\end{equation}
which is simply the tensor transformation from the original vertical flux $q^z$
when $z \ll r$ or $\theta_0 \approx \pi / 2$. Finally, the explicit form of the
flux is derived from the conservation laws, yielding
\begin{equation}
F(r)=-\frac{c^2 \dot{M}}{4 \pi \sqrt{-g / g_{\theta \theta}}} \frac{\partial_r
\Omega}{(E-\Omega L)^2} \int_{r_{\mathrm{ISCO}}}^r(E-\Omega L) \partial_r L d r.
\end{equation}

\bibliographystyle{unsrt}
\bibliography{xijipan}

\end{document}